%% file: example.tex
\documentclass[submission,copyright,creativecommons]{eptcs}
\providecommand{\event}{ICLP 2023} 

\usepackage{amsmath}
\usepackage{hyperref}
\usepackage{graphicx}
\usepackage{amsmath}
\usepackage{amssymb}
\usepackage{tabularx}
\usepackage{eurosym}
\usepackage{multirow}
\usepackage{inconsolata}
\usepackage{verbatim}
\usepackage{xspace}
\usepackage{xcolor}
\usepackage{fancyvrb}

\newcommand{\fc}{\xspace{\sc\small fogCutter}\xspace}
\newcommand{\cd}[1]{\texttt{\small #1}\xspace}

\DeclareMathOperator*{\argmax}{arg\!\max}

\usepackage{iftex}

\ifpdf
  \usepackage{underscore}         
  \usepackage[T1]{fontenc}        
\else
  \usepackage{breakurl}           
\fi

\title{Leasing the Cloud-Edge Continuum, à la Carte\thanks{Research partly funded by projects: \textit{Energy-aware management of software applications in Cloud-IoT ecosystems} (RIC2021PON\_A18), funded with ESF REACT-EU resources
by the \textit{Italian Ministry of University and Research} through the \textit{PON Ricerca e Innovazione 2014--20} and \textit{hOlistic Sustainable Management of distributed softWARE systems (OSMWARE)}, UNIPI PRA\_2022\_64, funded by the University of Pisa, Italy.}}
\author{Stefano Forti
\email{stefano.forti@unipi.it}
\institute{University of Pisa\\ Pisa, Italy}
\and
Ahmad Ibrahim
\email{a.ibrahim@bham.ac.uk}
\institute{University of Birmingham\\ Birmingham, United Kingdom}
\and
Antonio Brogi
\email{antonio.brogi@unipi.it}
\institute{University of Pisa\\ Pisa, Italy}
}

\def\titlerunning{Leasing the Cloud-Edge Continuum, à la Carte}
\def\authorrunning{S. Forti et al.}
\begin{document}
\maketitle

\begin{abstract}
Next-gen computing paradigms foresee deploying applications to virtualised resources along a continuum of Cloud-Edge nodes. Much literature has focused on how to place applications onto such resources so as to meet their requirements. To lease resources to application operators, infrastructure providers need to identify a portion of their Cloud-Edge assets to meet application operators' requirements. This article proposes a novel declarative resource selection strategy, prototyped in Prolog, to determine a suitable infrastructure portion that satisfies all requirements. The proposal is showcased over a lifelike scenario.
\end{abstract}

\input{src/introduction.tex}
\input{src/example.tex}
\input{src/fogcutter.tex}
\input{src/exampleretaken.tex}

\input{src/related.tex}

\input{src/conclusions.tex}

\bibliographystyle{eptcs}
\bibliography{biblio}

\end{document}

%% file: src/introduction.tex

\section{Introduction}
\label{sec:introduction}

Recent utility computing paradigms (e.g. Fog, Edge, Mist, Cloud-IoT computing) foresee deploying applications to virtualised resources along a continuum of Cloud-Edge computing, storage and networking resources. Much literature (surveyed, for instance, by~\cite{salaht2020overview} and~\cite{smolka2022evaluation}) has focussed on how to place (multi-service) applications onto such resources so as to meet their stringent functional and non-functional requirements (e.g. in terms of latency, bandwidth, security, deployment location, Quality of Service (QoS), energy sources, costs). 
Application placement has been proven an NP-hard problem (\cite{brogi2017qos}), featuring worst-case exp-time complexity of $O(N^S)$ for placing $S$ application services onto an infrastructure of $N$ nodes. Most of the proposed solutions (\textit{i}) take the perspective of application operators that need to deploy their application services, and (\textit{ii}) determine eligible placements by trying to match each service with any node in the infrastructure.

We tackle here a different, yet complementary, problem by (\textit{a}) \textit{taking the perspective of infrastructure providers} leasing Infrastructure-as-a-Service (IaaS) resources in the Cloud-Edge continuum to paying application operators that need to deploy their services, and by (\textit{b}) \textit{identifying a portion of the available Cloud-Edge infrastructure} that meets the requirements set by application operators and maximizes the expected profit that the infrastructure provider will get from leasing such infrastructure portion.

Said otherwise, we consider a setting in which:

\begin{enumerate}
    \item[(1)] Application operators request to lease a Cloud-Edge continuum \textit{portion}, featuring specific (hardware, software, security, network) \textit{resources} and suitably guaranteeing \textit{performance metrics} (e.g. sustainability, availability),
    \item[(2)] Infrastructure providers identify a portion that meets all the requirements of applications operators and that maximises their profit, by exploiting an automated (virtual) \textit{resource selection} mechanism,
    \item[(3)] Application operators can seamlessly deploy applications onto the leased infrastructure portion by employing any \textit{appplication placement strategy} of their choice.
\end{enumerate}

It is worth noting that such a change of perspective enables taming the exp-time complexity of the application placement problem by considering only a small portion of $n < N$ nodes in the available infrastructure.  For instance, our experiments in~\cite{fogbrainx} and~\cite{herrera2023continuous} show how placement execution times grow exponentially at increasing sizes of the considered infrastructure and, therefore, how reducing decision-making to smaller portions of such an infrastructure brings considerable speed-ups, viz. from 2$\times$ up to a $1000\times$ in realistic use cases. 

This article presents a declarative strategy to automate the aforementioned step (2)  by solving the following \textit{resource selection} problem:

\medskip
  \begin{minipage}{0.95\textwidth}
   \textit{Let $N$ be a set of heterogeneous nodes of a Cloud-Edge infrastructure managed by an infrastructure provider.
  Let $p: N \rightarrow \mathbb{R}$ be the function that defines the profit of the infrastructure provider for leasing a node.
  Let $R$ be a request of resources of an application operator.
  A solution to the considered resource selection problem is a portion $C = \{n_{1}, n_{2}, \dots n_{m}\} \subseteq N$ that guarantees all requirements in $R$ while maximising the infrastructure provider profit.
  Formally, given $N$, $p$, and $R$,  we aim at determining $C$ such that} 
  $$\argmax_{\substack{C\subseteq N\\ C \models R}} \sum_{n \in N} p(n).$$
  \end{minipage}

We present our novel open-source prototype\footnote{Available at \url{https://github.com/di-unipi-socc/fogCutter}} \fc, entirely written in Prolog. \fc allows infrastructure providers to describe the capabilities of their nodes and application operators to input their requests with a declarative syntax -- implying better readability, maintainability and extensibility. \fc will benefit both infrastructure providers and application operators. Infrastructure providers will identify suitable infrastructure portions to lease to maximise their profit. Application operators will reduce decision-making times for managing their distributed software by considering an infrastructure portion built \textit{à la carte}, specially based on their deployment needs.  

The rest of this article is organised as follows.
We describe a motivating scenario (Sect.~\ref{sec:example}) and illustrate the \fc Prolog prototype (Sect.~\ref{sec:methodology}), which we then use over the motivating scenario (Sect.~\ref{sec:exampleretaken}). Last, after discussing some related work (Sect.~\ref{sec:related}), we conclude pointing out to directions for future work (Sect.~\ref{sec:conclusions}).

%% file: src/example.tex
\section{Motivating Scenario}
\label{sec:example}

\begin{figure}[!h]
     \centering
     \includegraphics[width=0.95\textwidth]{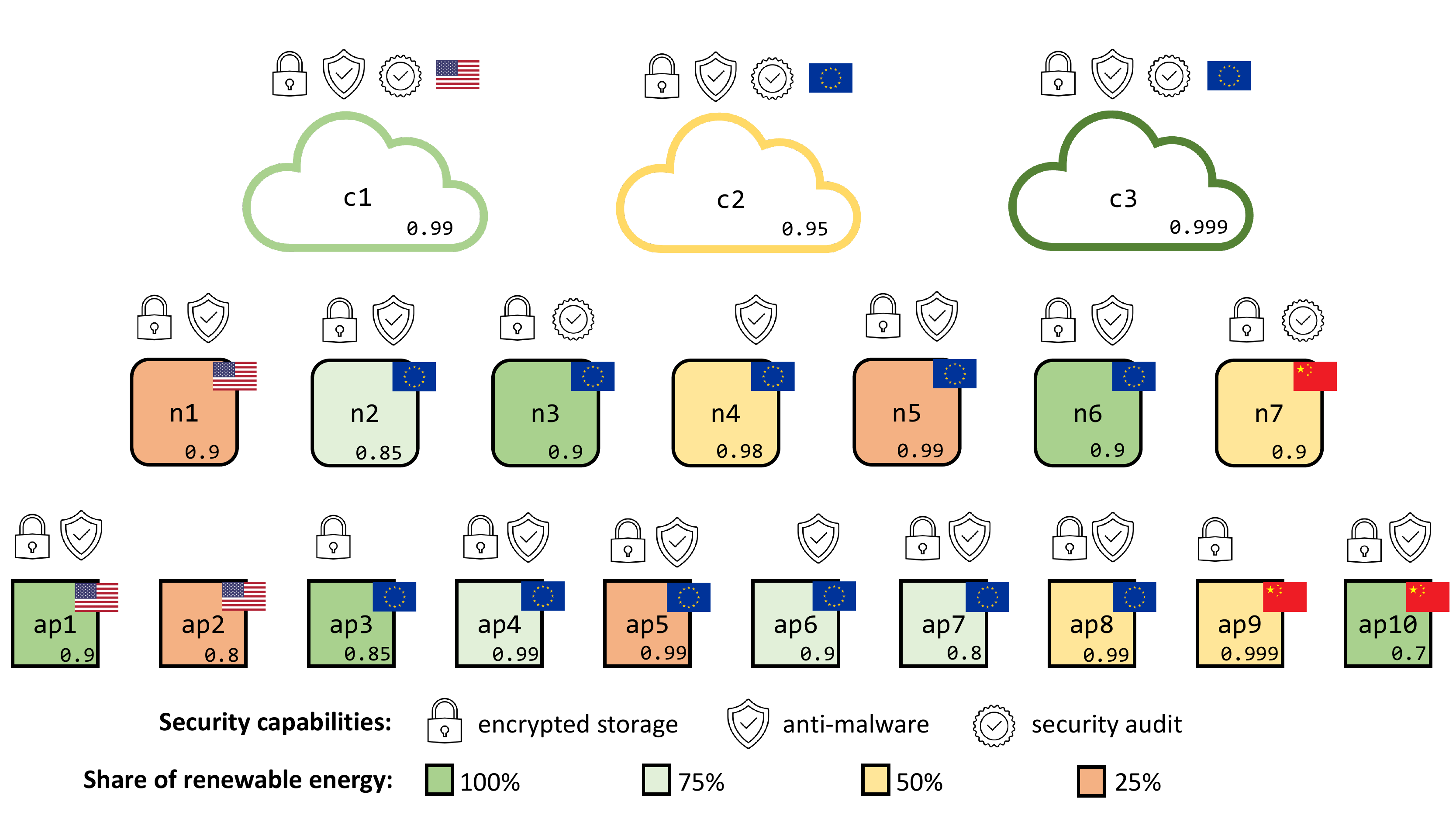}
     \caption{Available Cloud-Edge infrastructure.}
     \label{fig:infra}
 \end{figure}

Consider the Cloud-Edge continuum of Fig.~\ref{fig:infra}, provided by a single infrastructure operator across different geographical regions (viz. US, EU, China). The infrastructure is made of 20 nodes: 10 access points (APs) \cd{ap1}--\cd{ap10}, 7 more edge gateways \cd{n1}--\cd{n7} and 3 Cloud datacentres \cd{c1}--\cd{c3}. Assume the application operator connects to the network through \cd{ap3}.

 Each node in the available Cloud-Edge infrastructure of Fig. \ref{fig:infra} is characterised by its geographical location (represented as a flag), security capabilities (represented as icons), share of renewable energy it exploits to run (indicated by the colour of the node), and availability (reported in the bottom right side of each node). Finally, each node features a different amount of hardware and is associated to a profit value established by the infrastructure provider. Hardware and profit for each node are listed in Table~\ref{tab:hardware}.
 
As an example, the Cloud node \cd{c2} is located in the EU, features encrypted storage, anti-malware and security audit capabilities, is powered by 50\% of renewable energy, has an availability of 95\%, features 16 free hardware units, and -- when leased -- brings a profit of 7 \euro/hour to the infrastructure provider.

\begin{table}[!h]
\centering
\resizebox{0.8\textwidth}{!}{%
\begin{tabular}{|c|c|c|l|c|c|c|}
\cline{1-3} \cline{5-7}
\textbf{Node Id} & \textbf{Hardware Units} & \textbf{Profit} & { \ \ \ \ \ }  & \multicolumn{1}{l|}{\textbf{Node Id}} & \multicolumn{1}{l|}{\textbf{Hardware Units}} & \multicolumn{1}{l|}{\textbf{Profit}} \\ \cline{1-3} \cline{5-7} 
\cd{ap1}         & 2                       & 2.5             &  & \cd{n1}                               & 8                                            & 4                                    \\ \cline{1-3} \cline{5-7} 
\cd{ap2}         & 4                       & 1               &  & \cd{n2}                               & 12                                           & 5                                    \\ \cline{1-3} \cline{5-7} 
\cd{ap3}         & 3                       & 1.75            &  & \cd{n3}                               & 4                                            & 3                                    \\ \cline{1-3} \cline{5-7} 
\cd{ap4}         & 4                       & 3               &  & \cd{n4}                               & 12                                           & 4                                    \\ \cline{1-3} \cline{5-7} 
\cd{ap5}         & 2                       & 2.5             &  & \cd{n5}                               & 2                                            & 2.5                                  \\ \cline{1-3} \cline{5-7} 
\cd{ap6}         & 2                       & 1.5             &  & \cd{n6}                               & 8                                            & 4                                    \\ \cline{1-3} \cline{5-7} 
\cd{ap7}         & 4                       & 3               &  & \cd{n7}                               & 8                                            & 4                                    \\ \cline{1-3} \cline{5-7} 
\cd{ap8}         & 1                       & 2.25            &  & \cd{c1}                               & 24                                           & 9                                    \\ \cline{1-3} \cline{5-7} 
\cd{ap9}         & 3                       & 1.75            &  & \cd{c2}                               & 16                                           & 7                                    \\ \cline{1-3} \cline{5-7} 
\cd{ap10}        & 8                       & 4               &  & \cd{c3}                               & 12                                           & 6                                    \\ \cline{1-3} \cline{5-7} 
\end{tabular}%
}
\caption{Node hardware and profit.}
\label{tab:hardware}
\end{table}

As they are deployed at different geographical locations and interconnected with different (wired and wireless) technologies, end-to-end latency and bandwidth between nodes vary depending on the type of node and its location. 
For instance, on average, APs in the same location reach out to each other with latency of 10 ms and average symmetric bandwidth of 100 Mbps; those at different locations incur in a latency of 150 ms with a bandwidth of 25 Mbps. 
Besides, APs reach out edge nodes in their location with a latency of 60 ms, an upload bandwidth of 30 Mbps and a download bandwidth of 200 Mbps (viz. 200/30 Mbps). Last, APs connect to Cloud nodes at their location experiencing 130 ms latency and a 90/15 Mbps connection; those at a different location are reached out to with a latency of 200 ms and a 25/10 bandwidth. Data about the QoS of all other end-to-end links can be found online\footnote{Available at: \url{https://github.com/di-unipi-socc/fogCutter/blob/main/scenarios/smarttraffic.pl}}.

 An application operator aims at deploying a \textit{Smart Traffic} application onto a portion of the available infrastructure. Smart traffic management is a classical use case for Cloud-Edge applications (\cite{chabasnew2018,sehrapaving2022}), 
 having to collect and process data from various sensors (i.e. live video feed from traffic cameras, speed radar, air quality monitor sensor) and services (weather, date-time, location) to detect traffic congestion, issue speeding fines, dynamically adjust the price of tollgates and speed limits, control traffic signal duration, and send alerts and notifications. 
 
 Assume that the application operator accesses the network from node \cd{ap3}. To deploy her software artefacts, the application operator estimates she will need at least 20 free hardware units, reachable within 250 ms from node \cd{ap3} with a 10 Mbps bandwidth. Not to incur in too high management costs (i.e. time to decide where to place/migrate her application services at deployment and runtime, respectively), she would like her Cloud-Edge portion to include at most 4 nodes (\cd{ap3} included).

Receiving the above request, the infrastructure provider might wonder the following:

\begin{quote}
    \textbf{Q1.} \textit{Is there a portion of the available Cloud-Edge infrastructure that features at least 20 hardware units, and is made of at most four nodes with end-to-end connections featuring at least 10 Mbps bandwidth and at most 250 ms latency?}
\end{quote}

Before she can deploy her \textit{Smart Traffic} application, the application operator realises she also would need to meet GDPR constraints and deploy the software only to resources in the EU, provided with an antimalware and encrypted storage capabilities. She consequently updates her request to the infrastructure provider, who must answer the following question:

\begin{quote}
    \textbf{Q2.} \textit{Is there a portion of the available Cloud-Edge infrastructure that satisfies all requirements at} \textbf{Q1}\textit{, is fully located in the EU, and features antimalware and encrypted storage on all nodes?}
\end{quote}

Lastly, the application operator also determines that she wants the resources onto which to deploy her application to feature an overall availability of 85\% and to be likely to be powered by renewable energy at least 30\% of the time. Therefore, considering this new request, the infrastructure provider will look for an answer to the following question:

\begin{quote}
    \textbf{Q3.} \textit{Is there a portion of the available Cloud-Edge infrastructure that satisfies all requirements at} \textbf{Q1} \textit{and} \textbf{Q2} \textit{, and features 85\% availability and 30\% sustainability score?}
\end{quote}


%% file: src/fogcutter.tex
\section{Our Methodology}
\label{sec:methodology}

In this section, we first illustrate how \fc enables describing infrastructure capabilities and user requests with a simple and extensible model (Sect.~\ref{sec:model}). Then, we illustrate the logic program that, by processing such a model, determines solutions to the considered problem (Sect.~\ref{sec:cutter}). 

\subsection{Infrastructure and request model}
\label{sec:model}

Nodes are declared with a valid \cd{N} (e.g. their IP address, a symbolic name within the DNS) by means of facts like

\begin{Verbatim}[fontfamily=zi4, fontsize=\small, frame=single, framesep=1mm, framerule=0.1pt, rulecolor=\color{gray}]
node(N).
\end{Verbatim}

Each node is associated to the profit that its lease brings to the infrastructure provider, as in

\begin{Verbatim}[fontfamily=zi4, fontsize=\small, frame=single, framesep=1mm, framerule=0.1pt, rulecolor=\color{gray}]
profit(N, Profit).
\end{Verbatim}

After declaring a node, it is possible to declare its capabilities\footnote{Node capabilities can be obtained by tools targeting Cloud-Edge infrastructure monitoring, e.g. \cite{gaglianese2023assessing}.} through facts like

\begin{Verbatim}[fontfamily=zi4, fontsize=\small, frame=single, framesep=1mm, framerule=0.1pt, rulecolor=\color{gray}]
nodeCap(N, CapabilityType, FeaturedCapabilities).
\end{Verbatim}

\noindent that specifies the \cd{CapabilityType} and the currently \cd{FeaturedCapabilities} for such a type at node \cd{N}. 
For instance, node \cd{c1} of the motivating scenario of Sect. \ref{sec:example} is declared as in:

\begin{Verbatim}[fontfamily=zi4, fontsize=\small, frame=single, framesep=1mm, framerule=0.1pt, rulecolor=\color{gray}]
node(c1).
nodeCap(c1, hardware, 24).
nodeCap(c1, location, us).
nodeCap(c1, security, [antimalware, encryptedStorage, audit]).
nodeCap(c1, availability, 0.99).
nodeCap(c1, sustainability, 0.75).
\end{Verbatim}

Similarly, the capabilities of end-to-end links between nodes \cd{N} and \cd{M} can be declared as facts like

\begin{Verbatim}[fontfamily=zi4, fontsize=\small, frame=single, framesep=1mm, framerule=0.1pt, rulecolor=\color{gray}]
linkCap(N, M, Capability, FeaturedCapabilities).
\end{Verbatim}

As an example, the end-to-end link connecting nodes \cd{ap1} and \cd{n1} in the previous example can be expressed through the following facts:

\begin{Verbatim}[fontfamily=zi4, fontsize=\small, frame=single, framesep=1mm, framerule=0.1pt, rulecolor=\color{gray}]
linkCap(ap1,n1,latency,60).     linkCap(ap1,n1,bandwidth,30).
linkCap(n1,ap1,latency,60).     linkCap(n1,ap1,bandwidth,200).
\end{Verbatim}

We distinguish two kinds of capability types: \textit{local} and \textit{global}. Local capabilities are checked on a single resource, either a node or a link. Such capability types are declared through facts like

\begin{Verbatim}[fontfamily=zi4, fontsize=\small, frame=single, framesep=1mm, framerule=0.1pt, rulecolor=\color{gray}]
capType(Capability, Resource, ComparisonOperator).
\end{Verbatim}

\noindent that specify the \cd{Resource} for which such capability can be declared (viz. \cd{node}s or \cd{link}s), a \cd{Comparison-}\cd{Operator} to check whether the \cd{FeaturedCapabilities} at a node satisfy a request on such \cd{Capability}. For instance, \fc defines the following four local capabilities

\begin{Verbatim}[fontfamily=zi4, fontsize=\small, frame=single, framesep=1mm, framerule=0.1pt, rulecolor=\color{gray}]
capType(security,  node, supset).   capType(location,  node, member).
capType(latency,   link, smaller).  capType(bandwidth, link, greater).
\end{Verbatim}

\noindent that enable checking\footnote{For the sake of conciseness, comparison operators are not listed here. The interested reader can find them in the online project repository.} the set of security capabilities featured by a node against a set of requested capabilities, the location of a node against a list of permitted locations, and the latency and bandwidth featured by communication links against an upper and a lower bound, respectively.

Global capabilities are instead checked against a portion of the available infrastructure and also need to be aggregated. They can be declared through facts like

\begin{Verbatim}[fontfamily=zi4, fontsize=\small, frame=single, framesep=1mm, framerule=0.1pt, rulecolor=\color{gray}]
capType(CapabilityType, Resource, ComparisonOperator, AggregationOperator).
\end{Verbatim}

\noindent that also specify a suitable \cd{AggregationOperator} to compute them over a selected portion of the whole infrastructure. Again, \fc features the following default capabilities

\begin{Verbatim}[fontfamily=zi4, fontsize=\small, frame=single, framesep=1mm, framerule=0.1pt, rulecolor=\color{gray}]
capType(hardware,       node,   smaller,    sum).
capType(availability,   node,   smaller,    product).
capType(sustainability, node,   smaller,    product).
\end{Verbatim}

\noindent that correspond to the total amount of hardware of an infrastructure portion, and to the overall availability\footnote{We define \textit{availability} as the probability that a system is operational at a given time. It is obtained over a portion of an infrastructure by multiplying the availability of the single nodes in the portion.} and sustainability\footnote{We define \textit{sustainability} as the probability that a system is fully powered by renewable energy. It is obtained over a portion of an infrastructure by multiplying the sustainability of the single nodes in the portion. } featured by such a portion.
The model above is designed to be easily extensible by taking into account further node/link capabilities of arbitrary (local and global) types. 

\fc receives requests for Cloud-Edge portions through facts like

\begin{Verbatim}[fontfamily=zi4, fontsize=\small, frame=single, framesep=1mm, framerule=0.1pt, rulecolor=\color{gray}]
request(RequestId, SourceNode, MaxNodes, Requirements).
\end{Verbatim}

\noindent
where \cd{RequestId} is a unique identifier for the request, \cd{SourceNode} is the identifier of the user node from which the request is issued, \cd{MaxNodes} is the maximum number of nodes the user wants in her portion, and requirements is a list of pairs \cd{(CapabilityType, R)} that specifies which global/local capability is needed and the target amount or threshold \cd{R} that is requested by the user. For instance, the fact

\begin{Verbatim}[fontfamily=zi4, fontsize=\small, frame=single, framesep=1mm, framerule=0.1pt, rulecolor=\color{gray}]
request(req42, ap3, 3, [(hardware,20), 
                        (latency,250),(bandwidth,10),
                        (security,[antimalware, encryptedStorage]), (location,[eu]), 
                        (availability, 0.85), (sustainability, 0.3)]).
\end{Verbatim}

\noindent requires an infrastructure portion of at most 3 nodes in addition to \cd{ap3}, with at least 20 hardware units, reachable through end-to-end link featuring at most 250 ms latency and at least 10 Mbps bandwidth, featuring antimalware and encrypted storage capabilities, located in the EU, and with at least 85\% availability and 30\% sustainability.

\subsection{Resource Selection: a Logic Programming Strategy}
\label{sec:cutter}
Predicate \cd{portion/3} inputs a \cd{RequestId}, determines a \cd{Portion} of the available Cloud-Edge continuum that satisfies all the request's requirements and associates the \cd{Portion} with an estimate of the \cd{Profit} that the infrastructure provider will get from leasing such \cd{Portion}.
\begin{Verbatim}[fontfamily=zi4, fontsize=\small, frame=single, framesep=1mm, framerule=0.1pt, rulecolor=\color{gray}]
portion(RequestId, (Portion, Profit)) :-
    request(RequestId, N, MaxNodes, Reqs),      
    splitRequirements(Reqs, NodeReqs, LinkReqs, GlobalReqs),
    portion(N, MaxNodes, NodeReqs, LinkReqs, GlobalReqs, [N], Portion),
    portionProfit(Portion, Profit).
\end{Verbatim}
Predicate \cd{portion/3} first fetches the data (\cd{N}, \cd{MaxNodes}, \cd{Reqs}) of the input \cd{RequestId}.
Then predicate \cd{splitRequirements/4} simply partitions the list of request's requirements \cd{Reqs} into three lists, depending on whether they refer to local capabilities of nodes (\cd{NodeReqs}) or of links (\cd{LinkReqs}), or on whether they are global requirements on the overall infrastructure portion (\cd{GlobalReqs}).

\smallskip
\noindent
Predicate \cd{portion/7} determines a \cd{Portion} of the infrastructure that contains node \cd{N} and at most other \cd{MaxNodes} nodes and  satisfies all the 
\cd{NodeReqs}, \cd{LinkReqs} and \cd{GlobalReqs} requirements.
\begin{Verbatim}[fontfamily=zi4, fontsize=\small, frame=single, framesep=1mm, framerule=0.1pt, rulecolor=\color{gray}]
portion(N, MaxNodes, NodeReqs, LinkReqs, GlobalReqs, I, NewI) :-
    MaxNodes > 0,
    \+ satisfiesGlobalReqs(GlobalReqs, I),
    node(M), \+ member(M,[N|I]),
    satisfiesNodeReqs(M,NodeReqs),
    satisfiesLinkReqs(M,N,LinkReqs),
    NewMaxNodes is MaxNodes-1,
    portion(N, NewMaxNodes, NodeReqs, LinkReqs, GlobalReqs, [M|I], NewI).
portion(_, _, _, _, GlobalReqs, I, I) :- 
    satisfiesGlobalReqs(GlobalReqs, I).
\end{Verbatim}
Until all the global requirements \cd{GlobalReqs} are satisfied, predicate \cd{portion/7} recursively adds to the infrastructure portion a new node \cd{M} that satisfies all the \cd{NodeReqs} and \cd{LinkReqs} local requirements.

\smallskip
\noindent
Predicate \cd{satisfiesGlobalReqs/2} verifies whether the (non empty) infrastructure portion determined so far satisfies the list \cd{GlobalReqs} of global requirements.
\begin{Verbatim}[fontfamily=zi4, fontsize=\small, frame=single, framesep=1mm, framerule=0.1pt, rulecolor=\color{gray}]
satisfiesGlobalReqs(Rs, [I|Is]) :-  satisfies(Rs, [I|Is]).

satisfies([], _).
satisfies([(P,V,Op,Ag)|Rs], I) :-
    findall(Vm, (member(M,I),nodeCap(M,P,Vm)), Vs),
    aggregatedListValue(Vs,Ag,AV), compareValue(V,Op,AV),
    satisfies(Rs, I).

aggregatedListValue([V|Vs],Ag,AV):- 
    aggregatedListValue(Vs,Ag,AV2), aggregatedValue(V,Ag,AV2,AV).
\end{Verbatim}
Predicate \cd{satisfies/2} verifies that each global requirement \cd{(P,V,Op,Ag)} is satisfied by the infrastructure portion \cd{I} determined so far.
After fetching into a list \cd{Vs} all the values of property \cd{P} in the nodes \cd{M} in \cd{I}, \cd{satisfies/2} aggregates all the values in \cd{Vs} according to the indicated aggregator operator \cd{Ag} and compares the obtained aggregated value \cd{AV} with the value \cd{P} of the considered global requirement according to the indicated comparison operator \cd{Op}.
For instance, the aggregator operators \cd{sum} and \cd{product} and the comparison operators \cd{smaller} and \cd{greater} are defined as follows:
\begin{Verbatim}[fontfamily=zi4, fontsize=\small, frame=single, framesep=1mm, framerule=0.1pt, rulecolor=\color{gray}]
aggregatedListValue([],sum,0).
aggregatedListValue([],product,1).

aggregatedValue(X,sum,Y,Z) :- Z is X+Y.
aggregatedValue(X,product,Y,Z) :- Z is X*Y.

compareValue(X,smaller,Y) :- X < Y. 
compareValue(X,greater,Y) :- X > Y. 
\end{Verbatim}
Predicates \cd{satisfiesNodeReqs/2}  and \cd{satisfiesLinkReqs/2} verify that each local node and link requirement is satisfied if the new node \cd{M} is added to the  infrastructure portion \cd{I} determined so far.
\begin{Verbatim}[fontfamily=zi4, fontsize=\small, frame=single, framesep=1mm, framerule=0.1pt, rulecolor=\color{gray}]
satisfiesNodeReqs(M,[(P,V,Op)|Rs]) :- 
    nodeCap(M,P,Vm), compareValue(Vm,Op,V),
    satisfiesNodeReqs(M,Rs).
satisfiesNodeReqs(_,[]).
\end{Verbatim}
\begin{Verbatim}[fontfamily=zi4, fontsize=\small, frame=single, framesep=1mm, framerule=0.1pt, rulecolor=\color{gray}]
satisfiesLinkReqs(M,N,[(P,V,Op)|Rs]) :- 
    linkCap(M,N,P,Vmn), compareValue(Vmn,Op,V),
    satisfiesLinkReqs(M,N,Rs).
satisfiesLinkReqs(_,_,[]).
\end{Verbatim}
Predicate \cd{portionProfit/2} associates an infrastructure \cd{Portion} with an estimate of the \cd{Profit} that the infrastructure provider will get from leasing such \cd{Portion}.
The estimate of the overall \cd{Profit} can be simply obtained by summing the profit associated with each node in the determined \cd{Portion}.
\begin{Verbatim}[fontfamily=zi4, fontsize=\small, frame=single, framesep=1mm, framerule=0.1pt, rulecolor=\color{gray}]
portionProfit([N|Ns], Profit) :-
    profit(N, N_Profit),
    portionProfit(Ns, Ns_Profit),
    Profit is N_Profit + Ns_Profit.
portionProfit([], 0).
\end{Verbatim}
Finally, predicate \cd{fogCutter/2} inputs a \cd{RequestId} and determines the \cd{Portion} of the available Cloud-Edge continuum that satisfies all the request's requirements and that corresponds to the highest estimated \cd{Profit} for the infrastructure provider.
\begin{Verbatim}[fontfamily=zi4, fontsize=\small, frame=single, framesep=1mm, framerule=0.1pt, rulecolor=\color{gray}]
fogCutter(RequestId, (Portion,Profit)) :- 
    setof(X, portion(RequestId, X), CandidatePortions),
    bestPortion(CandidatePortions, (Portion,Profit)).

bestPortion(CandidatePortions, (Portion,Profit)) :- 
    member((Portion,Profit), CandidatePortions), 
    \+ (member((_,HigherProfit), CandidatePortions), Profit < HigherProfit). 
\end{Verbatim}

%% file: src/exampleretaken.tex
\section{{\sc fogCutter} in Action}
\label{sec:exampleretaken}


In this section, we exploit the \fc prototype to answer questions \textbf{Q1}, \textbf{Q2} and \textbf{Q3} of Sect.~\ref{sec:example}.

\medskip\noindent
We start by retaking question \textbf{Q1}:

\begin{quote}
    \textbf{Q1.} \textit{Is there a portion of the available Cloud-Edge infrastructure that features at least 20 hardware units, and is made of at most four nodes with end-to-end connections featuring at least 10 Mbps bandwidth and at most 250 ms latency?}
\end{quote}

The request of the application operator is denoted by a fact like:

\begin{Verbatim}[fontfamily=zi4, fontsize=\small, frame=single, framesep=1mm, framerule=0.1pt, rulecolor=\color{gray}]
request(req42, ap3, 3, [(hardware,20), (latency,250), (bandwidth,10)]).
\end{Verbatim}

\noindent The infrastructure provider can suitably query predicate \cd{fogCutter/2}, obtaining the following as a result:

\begin{Verbatim}[fontfamily=zi4, fontsize=\small, frame=single, framesep=1mm, framerule=0.1pt, rulecolor=\color{gray}]
?- fogCutter(req42,P).
P = ([ap3, ap8, c1, c2], 20.0) ;
false.
\end{Verbatim}

Therefore, \textit{there exist only one best candidate portion to lease to the application operator, associated to a profit of 20 \euro/h, and spanning node \cd{ap3}, \cd{ap8}, \cd{c1} and \cd{c2}. Note that there exist other 412 alternative solutions\footnote{We determined the number of alternative solutions by querying predicate \cd{setof(X,portion(RequestId, X),CandidatePortions)} and counting the found results.} that satisfy all the requirements of the application operator but feature a lower profit. } 

\medskip\noindent
We now consider question \textbf{Q2}:

\begin{quote}
    \textbf{Q2.} \textit{Is there a portion of the available Cloud-Edge infrastructure that satisfies all requirements at} \textbf{Q1}\textit{, is fully located in the EU, and features antimalware and encrypted storage on all nodes?}
\end{quote}

In this case, the request of the application operator is extended with security and location requirements, and can be expressed through the fact:

\begin{Verbatim}[fontfamily=zi4, fontsize=\small, frame=single, framesep=1mm, framerule=0.1pt, rulecolor=\color{gray}]
request(req42, ap3, 3, [(hardware,20), (latency,250), (bandwidth,10),
                        (security,[antimalware, encryptedStorage]), 
                        (location,[eu])]).
\end{Verbatim}

Querying again predicate \cd{fogCutter/2}, the infrastructure provider obtains:

\begin{Verbatim}[fontfamily=zi4, fontsize=\small, frame=single, framesep=1mm, framerule=0.1pt, rulecolor=\color{gray}]
?- fogCutter(req42,P).
P = ([ap3, ap4, c2, n2], 16.75) ;
P = ([ap3, ap7, c2, n2], 16.75) ;
false.
\end{Verbatim}

Therefore, \textit{this request can be satisfied by two different candidate portions of the available infrastructure that bring a maximum profit of 16.75 \euro/h, and span nodes \cd{ap3}, \cd{ap4}, \cd{c2} and \cd{n2}, or \cd{ap3}, \cd{ap7}, \cd{c2} and \cd{n2}, respectively. Also in this case, other 35 alternative portions exist that do not maximise the profit of the infrastructure provider.} 

\medskip\noindent
Last, we focus on answering question \textbf{Q3}:

\begin{quote}
    \textbf{Q3.} \textit{Is there a portion of the available Cloud-Edge infrastructure that satisfies all requirements at} \textbf{Q1} \textit{and} \textbf{Q2} \textit{, and features 85\% availability and 30\% sustainability score?}
\end{quote}

Here, the request of the application operator becomes:

\begin{Verbatim}[fontfamily=zi4, fontsize=\small, frame=single, framesep=1mm, framerule=0.1pt, rulecolor=\color{gray}]
request(req42, ap3, 3, [(hardware,20), (latency,250),(bandwidth,10),
                        (security,[antimalware, encryptedStorage]), 
                        (location,[eu]), 
                        (availability, 0.85), (sustainability, 0.3)]).
\end{Verbatim}

The query to \cd{fogCutter/2} from the infrastructure provider returns:

\begin{Verbatim}[fontfamily=zi4, fontsize=\small, frame=single, framesep=1mm, framerule=0.1pt, rulecolor=\color{gray}]
P = ([ap3, ap4, c2], 11.75).
\end{Verbatim}

Hence, \textit{the request can be satisfied by providing the application operator with the portion made from nodes \cd{ap3}, \cd{ap4} and \cd{c2}, with an associated profit of 11.75 \euro/h. For this last question, such a portion also represents the only eligible solution to the considered resource selection problem instance.}

\medskip\noindent
Table \ref{tab:results} recapitulates on the performance of the experiments \textcolor{black}{in our motivating scenario}, showing the number of Prolog inferences and the CPU time needed\footnote{Prolog inferences and CPU time are obtained by relying on SWI-Prolog's \cd{time/1} predicate. Experiments are obtained by running \fc in SWI-Prolog v. 8.4.3 (\cite{wielemaker:2011:tplp}) on a machine equipped with Apple Silicon M1 and 16GB of RAM.} to answer the queries. It also lists the number of optimal solutions (i.e. those that maximise profit and satisfy all requirements set by the application operator) and the number of eligible solutions (i.e. those that satisfy all requirements set by the application operator). It is worth noting that \fc determines an optimal solution almost instantaneously for all three questions.

\begin{table}[!h]
\centering
\resizebox{0.85\textwidth}{!}{%
\begin{tabular}{|c|c|c|c|c|}
\cline{1-5} 
{\textbf{Question}} & {\textbf{Prolog Inferences}} & 
{\textbf{CPU Time}} & {\textbf{Optimal Solution(s)}} & {\textbf{Eligible Solution(s)}} \\ 
\cline{1-5} 
\textbf{Q1}                             & 1,712,813                                & 0.186 s                                & 1                                               & 413                                                \\ \cline{1-5} 
\textbf{Q2}                             & 208,776                                  & 0.033 s                                  & 2                                               & 37                                                 \\ \cline{1-5} 
\textbf{Q3}                             & 343,393                                  & 0.048 s                                 & 1                                               & 1                                                  \\ \cline{1-5} 
\end{tabular}%
}
\caption{Performance of \fc over the motivating scenario.}
\label{tab:results}
\end{table}

 More in general, given the maximum number \cd{MaxNodes} of nodes requested for a portion, the worst-case time complexity for running \fc is $O(N^{\cd{MaxNodes}})$. 
%
 As mentioned before, determining valid placements of $S$ services onto $N$ nodes also features an exp-time complexity of $O(N^S)$. Setting, however, $\cd{MaxNodes} < S$ permits to quickly identify a portion of the infrastructure where to deploy applications and can dramatically reduce decision-making times related to placements. Indeed, as shown for instance in~\cite{fogbrainx} and~\cite{herrera2023continuous}, considering smaller problem instances can reduce execution times from minutes to milliseconds, with speed-ups from 2$\times$ up to a $1000\times$ in realistic use cases. Also, note that portions of the Cloud-Edge continuum can be pre-computed by the infrastructure provider over pre-defined request templates, and be immediately ready to be returned to application operators, when needed.



%% file: src/related.tex
\section{Related Work}
\label{sec:related}
  
Resource allocation is an important challenge in the Cloud-edge continuum. 
Lahmar and Boukadi \cite{lahmar2020resource} conducted a comprehensive survey of resource allocation challenges that should be considered while deploying an application over a Cloud-Edge continuum.
A similar survey by Shakarami et al. \cite{shakarami2022resource} classified existing resource allocation algorithms into five categories i.e., heuristic, framework, model, machine learning and game theory based, discussed their pros and cons and common performance metrics that are relevant for the application operators.
Faticanti et al. \cite{faticanti2020throughput} tackle a different problem which is to maximise the profit of the infrastructure provider by 
maximising the number of Fog applications in a batch that can be deployed over the Cloud-Edge continuum. To accomplish that their algorithm partition the application's components into two groups i.e., Cloud-only and Fog-only. Using throughput as QoS metrics, their algorithm tried to recommend appropriate regions for these components. 
 
 The role of the broker in the Cloud is also well investigated.
 Mei et al. \cite{mei2018profit} discuss the role and necessity of brokers in a Cloud-Edge continuum. According to them, the presence of a Cloud broker can lead to better pricing for the end user and at the same time, high utilisation of providers' resources. Cloud brokers can also earn profit in these transactions.  
 Wang et al. \cite{wang2017maximizing} discussed the idea that a Cloud broker
can maximise resource utilisation by differentiating between latency sensitive tasks (such as Fog applications) versus non-latency sensitive applications. They highlighted different strategies to deal with fairly pricing the two categories of applications for the broker that could make the offer attractive for a wide variety of users and providers.

The concept of IoT brokers was described in \cite{niyato2016smart, sciancalepore2017slice}.
Sciancalepore et al. \cite{sciancalepore2017slice} discussed a strategy that is aimed towards IoT brokers that can service different customers using IoT slices. In their approach, resources can be increased for a given slice as per demand in real-time.
A subscription based model for IoT providers on the Cloud-Edge continuum was described by 
Niyato et al. \cite{niyato2016smart}.  In their scheme, IoT owners and IoT data users have to subscribe to a service that enables data exchange data based on an agreed pricing model. They also described different strategies for IoT owners to bundle their offerings and make them attractive to the Cloud-Edge continuum users.

Slicing in a network is also thoroughly investigated although it is a slightly different problem for a different setting (i.e., telecom) \cite{wijethilaka2021survey}. 
IoT slicing is another related topic that deals with creating slices (partitions) of IoT resources in a Cloud-Edge continuum to satisfy different types of user requirements. Fernandez et al. \cite{fernandez2019enabling} proposed a concept of an IoT orchestration service that could satisfy a customer request, especially on a resource constraint platform using IoT slices. Similarly, Hwang et al. \cite{hwang2021iot} described a central service to manage incoming requests (tasks), partition the infrastructure into different slices of the IoT sensors and distribute the tasks to different IoT slices to answer the query.
Experimental results conducted by Luthra et al. \cite{luthra2021tcep} show that different operator placement algorithms are needed to suit the dynamic environment and conflicting QoS demands.


Our approach is different from the existing approaches as we try to solve a \textit{resource selection} problem marginally addressed before. Our approach allows a declarative and extensible definition of parameters which can be adopted as per application operator and infrastructure operator requirements.

%% file: src/conclusions.tex
\section{Concluding Remarks}
\label{sec:conclusions}
The problem of selecting an infrastructure portion matching the requirements set by an application operator is challenging in the 
Cloud-Edge continuum due to the large number of assets with heterogeneous capabilities. 
The problem is important both for application operators, who need to suitably deploy their applications, and for infrastructure providers, who wish to maximize their profit in leasing infrastructure resources to application operators.

In this article, we have presented a declarative resource selection strategy for the Cloud-Edge continuum, implemented in the open source prototype \fc.
Using a motivating example, we have shown how \fc can be successfully exploited to reply to different requests of customers (viz. application operators).
The main advantages of the declarative solution implemented by \fc are its readability and ease of extensibility to include new types of requirements. Indeed, it is straightforward to include further local (e.g. availability of IoT devices or specialised hardware, link jitter) and global requirements (e.g. QoS-assurance, nodes' trust) by simply programming new \cd{capType} predicates and the associated comparison and aggregation operators. 
Directions for future work include:

\begin{itemize}
    \item returning partial solutions (when not all requirements can be met), considering partial allocations of Cloud-Edge resources, and allowing customers to declare 
    hard and soft constraints in requests,
    \item performing a thorough scalability assessment at varying sizes of the input Cloud-Edge infrastructure and number of considered requirements for the portion to be determined, also extending \fc with heuristic strategies, and
    \item assessing the tool with real data (possibly against alternative solutions, e.g. constraint-based) in simulated, laboratory or testbed settings, also in pipeline with placement strategies to better estimate the obtainable speed-ups.
\end{itemize}

